\documentclass[aps,
citeautoscript,
floatfix,twocolumn,superscriptaddress]{revtex4-1}
\usepackage{graphicx,subfigure,epsfig,verbatim,psfrag,amsmath,amssymb,color}
\usepackage{dcolumn}
\usepackage{bm}
\usepackage{hyperref}
\usepackage{natbib}

\begin{document}

\title{Criticality and scaling corrections for two-dimensional Heisenberg models in plaquette patterns with strong and weak couplings}

\author{Xiaoxue Ran}
\affiliation{State Key Laboratory of Optoelectronic Materials and
Technologies, School of Physics, Sun Yat-Sen University, Guangzhou 510275, China}
\author{ Nvsen Ma}
\email{nvsenma@iphy.ac.cn}
\affiliation{Beijing National Laboratory of Condensed Matter Physics and Institute of Physics,
Chinese Academy of Sciences, Beijing 100190, China}
\author{ Dao-Xin Yao}
\email{yaodaox@mail.sysu.edu.cn}
\affiliation{State Key Laboratory of Optoelectronic Materials and
Technologies, School of Physics, Sun Yat-sen University, Guangzhou 510275, China}

\begin{abstract}
We use the stochastic series expansion quantum Monte Carlo method to study the Heisenberg models on the square lattice with strong and weak couplings in the form of three different plaquette arrangements known as checkerboard models C$2\times2$, C$2\times4$, and C$4\times4$. The $a\times b$ here stands for the shape of the plaquette consisting of spins connected by strong couplings. Through detailed analysis of a finite-size scaling study, the critical point of the C$2\times2$ model is improved as $g_{c}=0.548524(3)$ compared with previous studies where $g$ is the ratio of weak and strong couplings in the models. For C$2\times4$ and C$4\times4$ we give $g_{c}=0.456978(2)$ and $0.314451(3)$. We also study the critical exponents $\nu$ and $\eta$ and the universal property of the Binder ratio to give further evidence that all quantum phase transitions in these three models are in the three-dimensional O(3) universality class. Furthermore, our fitting results show the importance of effective corrections in the scaling study of these models.

\end{abstract}
\maketitle
\section{Introduction}\label{sec:1}
The $S=1/2$ Heisenberg antiferromagnetic model with different interactions\cite{overall,Murg2009} has always been a very interesting topic in both theoretical and experimental fields because of its rich ground states and
close relations to cuprate superconductors\cite{Horsch1988,cup91,Keimer1992,Lee2006}, Bose-Einstein condensation of magnons\cite{bec08,Rakhimov2012}, etc. One of the best studied two-dimensional (2D) Heisenberg models is
the dimerized model\cite{Matsumoto2001,andersbook,Jiang2012,accurate} with inter- and intradimer antiferromagnetic couplings on the square lattice, which bring in quantum fluctuations to destroy the N$\acute{e}$el ground state and
make the model undergo a quantum phase transition (QPT)\cite{always} from antiferromagnetic (AFM) to quantum paramagnetic (QPM)\cite{cubic,accurate,different,wangl06,Syljuasen2006,gc,yao10}. Based on field analysis mapping
to a nonlinear $\sigma$ model, this QPT belongs to the three-dimensional (3D) O(3) universality class\cite{3dmap}, which is also proved by several separate numerical results with high accuracy\cite{different, ma18, suwa15,
wangl06}.

Apart from those well studied dimerized models, a QPT from AFM to QPM can also be realized by introducing strong and weak couplings which favor the formation of singlets in quadrumerized or other patterns, which connect more spins as long as there are an even number of strong couplings in the unit\cite{Koga1999,wangl06,Mambrini2006,different, troyer08}. These patterns are referred to as the checkerboard patterns, which were proposed to explain the
experiments of real-space structures observed in Bi$_{2}$Sr$_{2}$CaCu$_{2}$O$_{8+\delta}$ and Ca$_{2-x}$Na$_{x}$CuO$_{2}$Cl$_{2}$\cite{experiment1, experiment2,Shen2005}. The quadrumerized Heisenberg model on the square lattice with
$2\times2$ spins connected by stronger couplings can also be very helpful in the study of the  Shastry-Sutherland model\cite{ssmodel, wessel02}, which explains the critical properties of
SrCu$_{2}$(BO$_{3}$)$_{2}$\cite{ssexperiment}. Recently there has been a very popular discussion about the order of QPT in the Shastry-Sutherland model\cite{ssdqcp,bowen}. However, except for the quadrumerized Heisenberg
model, no numerical study has ever been done on plaquette models in which larger numbers of spins have been connected by strong couplings. Even for the quadrumerized model, the previous best estimate for the critical point is
$g_{c}=1.8230(2)$, with $g$ being the ratio of strong and weak couplings in the system\cite{wenzel09}, whose accuracy is at least one order of magnitude larger than that of the dimerized model (e.g., $g_{c}=1.90951(1)$ in the columnar dimerized model (CDM)\cite{ma18} ) or the classical 3D Heisenberg model\cite{classical}. We note that the coupling ratio $g$ in this work is weak couplings divided by strong couplings while in dimerized models it is
reciprocal as mentioned above. Besides, a recent work concerning QPT from AFM to QPM  shows that different local symmetries may bring in different critical corrections at QPTs\cite{ma18}. It has answered a long standing issue
that the quantum Monte Carlo (QMC) simulation results of critical exponents in a staggered dimerized model (SDM) are not the standard O(3) values\cite{different,cubic}. In their work they compared the critical exponents and
correction forms of the SDM with those of the CDM to show that they belong to the same universality class with different corrections. However, the previous work that had claimed to find different exponents in the SDM also compared it
with the quadrumerized Heisenberg model, whose correction form has not been carefully studied yet.

\begin{figure}
 \label{fig1}
  \centering
  \includegraphics{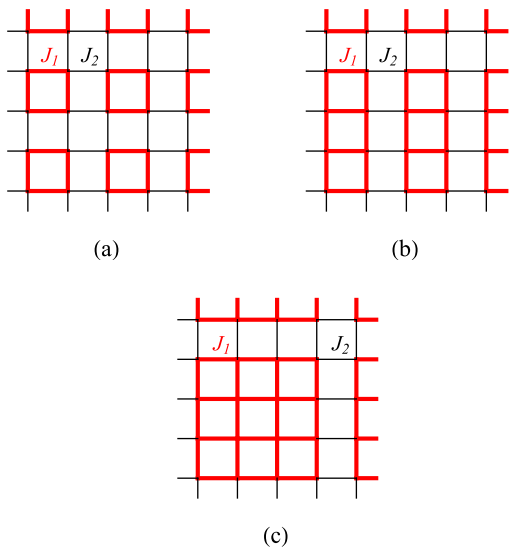}
  \caption{(Color online) The antiferromagnetic Heisenberg model defined in Eq.(\ref{eq1}) with $J_{1}$ (red thick bonds) arranged in units (a) $2\times2$, (b) $2\times4$, and (c) $4\times4$ on a square lattice. For each model,
  all units are connected by weak couplings $J_{2}$ (black thin bonds) to form the checkerboard pattern.}
\end{figure}

In this paper we study a series of plaquette antiferromagnetic Heisenberg models on a square lattice with the Hamiltonian
\begin{equation}
\label{eq1}
H=J_{1}\sum_{\langle i,j\rangle}{\bf S}_{i}\cdot{\bf S}_{j}+J_{2}\sum_{\langle i,j\rangle'}{\bf S}_{i}\cdot{\bf S}_{j},
\end{equation}
where ${\bf S}_{i}$ denotes an $S=1/2$ spin operator at lattice sites $i$, and $\langle i,j\rangle$ and $\langle i,j\rangle'$ are the nearest-neighbor sites connected by corresponding coupling strengths, which are represented by
strong couplings $J_1$ and weak couplings $J_2$.
According to different checkerboard patterns of the arrangements of $J_{1}$ and $J_2$ shown in Fig.~\ref{fig1}, we refer to these plaquette models as the C$2\times2$, C$2\times4$, and C$4\times4$ models. We set $J_{1}=1$ and define the ratio of weak and strong couplings to be $g=J_{2}/J_{1}$. When $g=1$, the model becomes an isotropic Heisenberg plane which has an antiferromagnetic ground state with long-range order. When $g=0$, the ground state turns into a disordered phase with no magnetism. It is a product state of singlets that differs in different plaquette models\cite{Xu2019}. In
this case, for $0<g<1$ there is a critical point $g_{c}$ at zero temperature where a QPT from AFM to QPM would happen. This QPT is in the O(3) universality according to the nonlinear $\sigma$ mapping analysis
class\cite{3dmap}. We use the stochastic series expansion (SSE) QMC method\cite{andersbook} and a finite-size scaling (FSS) study to estimate the critical points and exponents in the thermodynamical limit for all three models.

There are mainly two purposes to studying these three plaquette models. The first one is that we want to obtain the critical point of the C$2\times2$ (quadrumerized) model with higher accuracy by large scale QMC calculation and a
detailed FSS study on several different variables. The final result for the C$2\times2$ model is $g_{c}=0.548524(3)$, which is improved quite obviously compared with $g_{c}=0.54854(1)$ as the previous best estimate. Except for
being helpful to the exploration of finite-$\emph{T}$ quantum criticality by minimizing the quantum regime in the C$2\times2$ model itself, the increase of the statistical accuracy here would be useful to study the related
models (i.e., Shastry-Sutherland model) too\cite{wessel02}. We also obtain $g_{c}=0.456978(2)$ and  $g_{c}=0.314451(3)$ for the C$2\times4$  and C$4\times4$ models, respectively, for the first time. This could be very
necessary in future study of a certain material with this kind of real-space structure in the experiment. These results of critical points offer a very good benchmark for further tests or development of numerical
techniques of FFS methods as well. Second, from the FSS of criticality of the C$2\times2$ model, we can determine its correction form and correction exponent to compare with those of the SDM as a complement to the comparison in
Ref.~[\onlinecite{ma18}]. The nonmonotonic scaling behavior is not found in the criticality with critical exponents in the O(3) universality class in any of these three models, which is similar to the CDM.
Thus, our results give more examples that having local symmetry lacking cubic couplings which can bring in corrections not present in the standard O(3) class. The scaling correction forms and correction exponents in
different models are the same as each other, which implies that the different local $Z_{4}$ (C$2\times2$  and C$4\times4$) or $Z_{2}$ (C$2\times4$) symmetries do not result in any difference. These results could be helpful in
understanding the QPTs with irrelevant field.

In addition to the main purposes above, the standard O(3) value $1/\nu=1.4061(7)$ is chosen for all three models from the 3D classical Heisenberg model\cite{classical} and $\chi^{2}/d.o.f$ ($\chi^{2}$ per degree of freedom) close to 1 for all fits. Through these scalings, all critical points obtained from different quantities agree with each other in one system, which offers computational evidence to confirm that QPTs in all three models belong to the O(3) universality class. Besides $\nu$, we also compare the anomalous dimension $\eta$ and the dimensionless Binder ratio at critical points as further evidence to prove the predicted universality class.

The rest of the paper is organized as follows. In Sec.~\ref{sec:2} we introduce the physical quantities calculated in this work and the finite-size scaling method that we utilize to analyze data from simulations. In Sec.~\ref{sec:3} simulation and FFS results of criticalities for all three models are presented with detailed analysis. In the end we give a brief summary and discussion in Sec.~\ref{sec:4}.

\section{Observables and Finite-size scaling}\label{sec:2}

We use the SSE QMC simulation method with an operator-loop updating algorithm to study all plaquette models in our work. This computing method is based on sampling of the diagonal elements of the Boltzmann operator $exp(-\beta H)$, with $\beta$ being the inverse temperature. In order to rule out the effect of temperature on the scaling function near the quantum critical point, $\beta$ is always chosen as $\beta\sim L^{z}$. The QPTs in the plaquette models studied here are believed to be in accordance with O(3) behavior so that $z=1$\cite{Troyer1997} in these models; therefore we consider $\beta=L$ in our study with simulated system size $L$ up to $160$. In all calculations we
use $~10^{4}$ Monte Carlo samplings to obtain average values of the observables.
\subsection{Observables}
In order to study the criticality of the certain spin model, we chose to measure several important physical quantities in our work. The first one is the Binder ratio defined as
\begin{equation}
R_{2}=\frac{\langle (m_{s}^{z})^{4}\rangle} {\langle(m_{s}^{z})^{2} \rangle^{2}},
\label{binder}
\end{equation}
where
\begin{equation}
m_{s}^{z}=\frac{1}{N}\sum_{i}^{N}S_{i}^{z}(-1)^{x_{i}+y_{i}},
\label{subm}
\end{equation}
where $N=L\times L$ is the total number of spins on the square lattice and $(x_{i},y_{i})$ are the coordinates of the corresponding spin $S_{i}$. The Binder ratio is dimensionless and universal regardless of the detailed
structures and couplings of the model. However it does depend on the boundary conditions and effective aspect ratios from previous studies\cite{Hasenbusch2001,Selke2007,Selke2009,Qian2016}. Here we use periodic boundary conditions on these three models and the effective aspect ratio of time-space is related to the critical spin-wave velocity.

Another quantity is the uniform susceptibility
\begin{equation}
\chi_{u}=\chi(0,0)=\frac{\beta}{N}\left\langle\left(\sum_{i=1}^{N}S_{i}^{z}\right)^{2}\right\rangle,
\label{uniform}
\end{equation}
whose scaling form at $g_{c}$ is $\chi_{u}\sim L^{z-d}$, giving $\chi_{u}\sim L^{-1}$ and $\chi_{u}L$ to be dimensionless in our case.

The last physical observable calculated in our work is the spin stiffness $\rho_{s}$. The stiffness $\rho$ is covered in the calculation
\begin{equation}
\delta f=\frac{1}{2}\rho(\nabla\theta)^{2}=\frac{1}{2}\rho(\Phi/L)^{2}
\label{fieldrho}
\end{equation}
in the continuum field theory with $f$ being the density of free energy, $\Phi$ the boundary twist, and $\theta$ the order parameter field. In the Heisenberg model, $\rho_{s}$ is the spin stiffness determined by the twist $\Phi$ directly
to the Hamiltonian, which in the SSE procedure can be obtained through the calculation
\begin{equation}
\rho_{s}^{a}=\frac{3}{2\beta N}\langle (N_{a}^{+}-N_{a}^{-})^{2}\rangle,
\label{spinstiff}
\end{equation}
where $N_{a}^{+}$ ($N_{a}^{-}$) represent the total number of $S^{+}_{i}S^{-}_{j}$ ($S^{-}_{i}S^{+}_{j}$) operators in the sampling along the $a$ ($x$ or $y$) direction of the square lattice.  When the system is isotropic, the lattice $\rho_{s}^{x}$ is the same as $\rho_{s}^{y}$, while for the anisotropic system they are different. So in the C$2\times2$ and C$4\times4$ models we only calculate $\rho_{s}=(\rho_{s}^{x}+\rho_{s}^{y})/2$, and for the C$2\times4$
model both $\rho_{s}^{x}$ and $\rho_{s}^{y}$ are recorded separately. However they all have the same scaling form at the critical point as $\rho_{s}\sim L^{2-d-z}$, with $\rho_{s}\sim L^{-1}$ in our models, which means that
$\rho_{s}L$ is a size-independent dimensionless quantity.

\subsection{Finite-size scaling }
After all the mean observable values mentioned above are obtained from the simulations, we need to deal with all these data using the finite-size scaling study method to estimate the critical properties in the
thermodynamical limit\cite{andersbook,Jiang2012,accurate,cubic,different,wangl06,ma18,troyer08,Shao2016,Liu2018}. From the renormalization group theory we know that the physical quantity $Q$ near its critical point obeys

\begin{equation}
Q(g,L)=L^{\kappa/\nu}f(\delta L^{1/\nu},\lambda_{1}L^{-\omega_{1}},\lambda_{2}L^{-\omega_{2}},\dots),
\label{ffs1}
\end{equation}
where $\kappa$ is the critical exponent of $Q$, $\nu$ is the correlation length exponent, and $\delta=g-g_{c}$.  The set $\{\lambda_{i}\}$ refers to all irrelevant fields with their correction exponents  $\{\omega_{i}\}$,
which are arranged as $\omega_{i+1}>\omega_{i}$. Usually at most one irrelevant field is supposed to be considered in the FSS analysis, but there are still some special cases where more than one field is
necessary\cite{ma18}. Here we start with one correction exponent to the first order of the dimensionless quantities ($\kappa=0$) so that Eq.(\ref{ffs1}) can be written as
 \begin{equation}
Q(g,L)=f_{Q}^{(0)}(\delta L^{1/\nu})+L^{-\omega_{1}}f_{Q}^{(1)}(\delta L^{1/\nu}),
\label{leading}
\end{equation}
in which $L^{-\omega_{1}}$ is regarded as  a deviation value of the theoretical scaling function $f_{Q}$ near the critical point. Ignoring irrelevant items, the dimensionless quantity $Q(g,L)$ does not depend on the size of the system at the critical point $g_{c}$ because $g=0$ then. Thus,  $Q(g,L)$ values for different sizes cross at the critical point in this simplified situation, but here we need to take irrelevant items into consideration and $Q(g,L)$ values for different
sizes would cross at $g_{c}(L)$, which is near to the real $g_{c}$ with a correction to the order $L^{-\omega_{1}}$. For two different simulated sizes $L$ and $L^{'}$, using Eq.(\ref{leading}), we have
\begin{equation}
\begin{split}
f_{Q}^{(0)}(g^{*}L^{1/\nu})+L^{-\omega_{1}}f_{Q}^{(1)}(g^{*}L^{1/\nu})\\
=f_{Q}^{(0)}(g^{*}L^{'1/\nu})+L^{'-\omega_{1}}f_{Q}^{(1)}(g^{*}L^{'1/\nu})
\end{split}
\label{two}
\end{equation}
at the cross point $g_{c}(L)$, with $g^{*}=g_{c}(L)-g_{c}$. Expanding $f_{Q}^{(0)}$ and $f_{Q}^{(1)}$  to the first order of $L^{-\omega_{1}}$ with $L^{'}=bL$ we can get
\begin{equation}
g^{*}=\frac{f_{Q}^{(1)}(0)}{f_{Q}^{(0)'}(0)}\frac{b^{-\omega_{1}}(b^{\omega_{1}}-1)}{b^{1/\nu}-1}L^{-\omega_{1}-1/\nu},
\label{crossx}
\end{equation}
which is more easily understood as
\begin{equation}
g_{c}(L)=g_{c}(\infty)+\frac{f_{Q}^{1}(0)}{f_{Q}^{(0)'}(0)}\frac{b^{-\omega_{1}}(b^{\omega_{1}}-1)}{b^{1/\nu}-1}L^{-\omega_{1}-1/\nu}.
\label{crossx_new}
\end{equation}
Insert Eq.~(\ref{crossx}) into Eq.~(\ref{leading}) and again expand $f_{Q}^{(0)}$ and $f_{Q}^{(1)}$  to the first order of $L^{-\omega_{1}}$, then we have
\begin{equation}
Q(L)=Q_{\infty}(g_{c})+\frac{b^{-\omega_{1}}(1-b^{1/\nu+\omega_{1}})f_{Q}^{(1)(0)}}{b^{1/\nu}-1}L^{-\omega_{1}}.
\label{crossy}
\end{equation}
 Besides, with the definition of
 \begin{equation}
\frac{1}{\nu(L)}=\frac{1}{ln(b)}\left(ln\frac{S(L^{'})}{S(L)}\right),
\label{defnu}
\end{equation}
where
 \begin{equation}
S(L)=\frac{dQ(g,L)}{dg}\bigg|_{g=g_{c}(L)},
\label{defines}
\end{equation}
we can also obtain the scaling of critical exponent $\nu$ combining Eqs.(\ref{crossx_new}) and (\ref{defnu}) as
 \begin{equation}
 \frac{1}{\nu(L)}=\frac{1}{\nu}+aL^{-\omega}
\label{scalenu}
\end{equation}
with a free parameter $a$.
For simplicity the scaling forms of the coordinates of crossing points $(g_{c}(L),Q_{c}(L))$ are written as
\begin{align}
g_{c}(L) &=g_{c}(\infty)+bL^{-\omega-1/\nu}\label{scaling1},\\
Q_{c}(L) &=Q_{c}(\infty)+cL^{-\omega},
\label{scaling}
\end{align}
with $b$ and $c$ to be fitted as free parameters. From Eqs.(\ref{scaling1}) and (\ref{scalenu}), we know that by using crossing points from the $g$ dependence of the dimensionless quantity for two sizes $(L and bL)$, the extrapolation
value when $L\rightarrow\infty$ can give the quantum critical point $g_{c}$ and the critical exponent $\nu$ at the thermodynamical limit. In our work, we use $b=2$ in obtaining all crossing points for different models.

\section{Simulation results and data analysis}\label{sec:3}
\begin{figure}[!]
  \centering
  \includegraphics{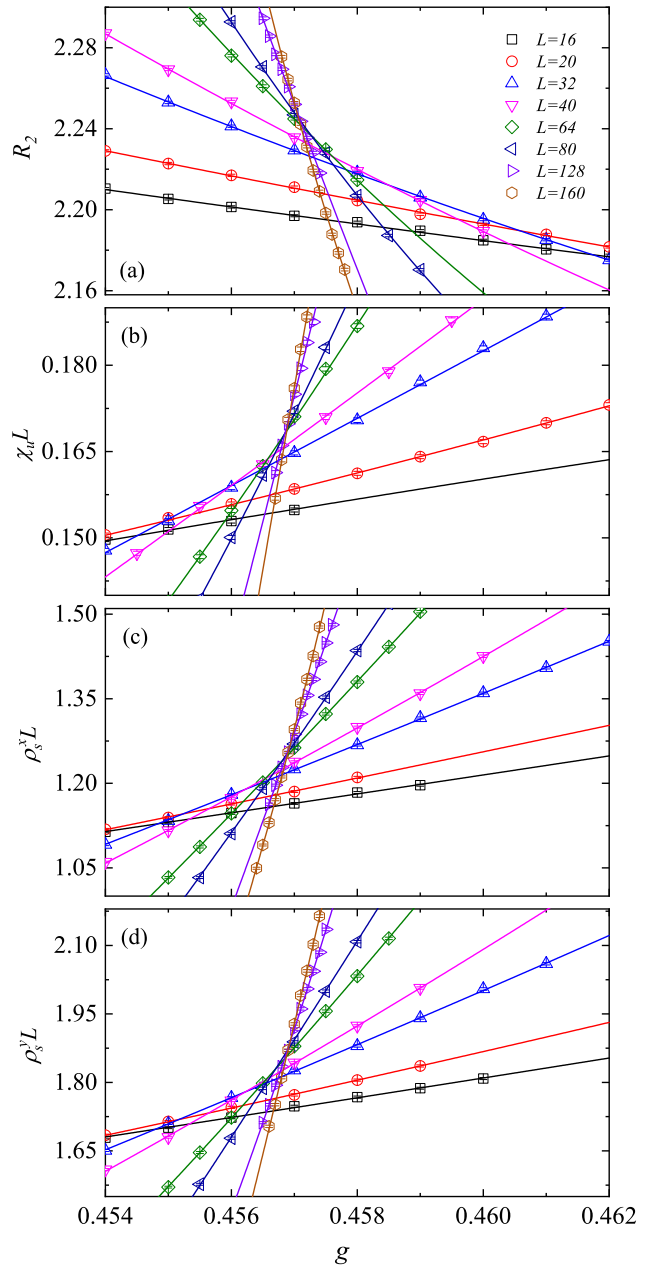}\\
  \caption{(Color online) The Binder ratio (a), uniform susceptibility multiplied by $L$ (b), and spin stiffness in the $x$ (c) and $y$ (d) directions multiplied by $L$ versus the coupling ratio $g$. All data points are connected by polynomial fitted curves to the third order with $\chi^{2}/d.o.f$ close to 1. This means that the ranges of $g$ are chosen correctly as being close enough to $g_{c}$ and for larger sizes the fitting range is smaller. We also adjust the display range of all subgraphs to make the shift of the crossings more clear.}
 \label{fig2}
\end{figure}

We performed the SSE QMC simulations on the C$2\times2$, C$2\times4$, and C$4\times4$ models and obtained the average values of all the observables $R_{2}$,
$\chi_{u}$, and $\rho_{s}$ (or $\rho_{s}^{x}$ and $\rho_{s}^{y}$ especially for C$2\times4$). One example of the simulation results for C$2\times4$ is illustrated in Fig.~\ref{fig2} to show the crossings of different sizes
for the four dimensionless quantities  $R_{2}$, $\chi_{u}L$ , $\rho_{s}^{x}L$, and $\rho_{s}^{y}L$.  Similar figures can also be obtained from the SSE data for the C$2\times2$ and C$4\times4$ models. The obvious shift of crossings from
different sizes implies that it is necessary to take the correction into account in the scaling analysis.

\subsection{Critical points and corrections}
\begin{figure}
  \includegraphics{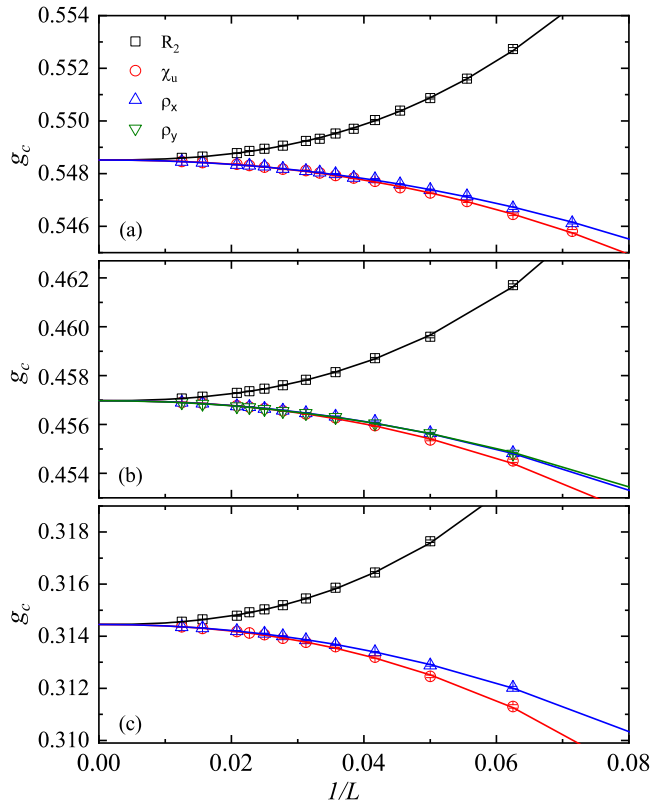}\\
  \caption{(Color online) Size dependence of  $g_{c}(L)$ of all $(L,2L)$ crossings from the $g$ dependence of $R_{2}$, $\chi_{u}L$ and $\rho_{s}L$ in the (a) C$2\times2$ , (b) C$2\times 4$, and (c) C$4\times4$ models. The curves
  stand for the fitting function in Eq.~(\ref{scaling1}) with the same $g_{c}(\infty)$ in each model and the fixed value $1/\nu=1.406$. The joint fits give $g_{c}=0.548524(3)$ and $\omega=1.12(2)$, $0.82(3)$, and $0.67(1)$ for $R_{2}$,
  $\chi_{u}L$, and $\rho_{s}L$, respectively, in panel (a); $g_{c}=0.456978(2)$ and $\omega=1.06(1)$, $0.83(1)$, $0.692(8)$, and $0.62(1)$ for $R_{2}$, $\chi_{u}L$, $\rho_{s}^{x}L$, and $\rho_{s}^{y}L$, respectively, in panel (b); and
  $g_{c}=0.314451(3)$, $\omega=1.04(1)$ and $0.84(1)$ and $0.66(1)$ for $R_{2}$, $\chi_{u}L$, and $\rho_{s}L$, correspondingly in panel (c). All fits give the estimates of $g_{c}$ and $\omega$ with $\chi^{2}/d.o.f$ close to 1. }
  \label{fig3}
\end{figure}

After all crossing points are extracted from the raw data we use the finite-size scaling method to estimate the critical points of our models. By fitting all the data points in Fig.~\ref{fig3} using the function in Eq.~(\ref{scaling1}) separately for each quantity we obtain all the $g_{c}$ results shown in Table~\ref{tab1}. In all the fits we use the standard O(3) value $1/\nu=1.406$ with one correction exponent $\omega$ to the first order. For the C$2\times2$ model, $g_{c}$ from each quantity is the same considering one error bar and agrees with the former result $g_{c}=0.54854(1)$. It is also true for the other two models with $\chi^{2}/d.o.f$ close to 1, implying the credibility of the fits. These results give further evidence to show that plaquette models with different checkerboard patterns all belong to the O(3) universality class.
\begin{table}[b]
\caption{
The finite-size scaling results of the critical point $g_{c}$ using different quantities for the C$2\times2$ (top), C$2\times 4$ (middle), and C$4\times 4$(bottom) models. Here we fit all the data points in Fig.~\ref{fig3} using
scaling formula Eq.~(\ref{scaling1}) with $1/\nu=1.406$.
}
\begin{ruledtabular}
\begin{tabular}{ldcc}
\textrm{}&
\multicolumn{1}{c}\textrm{$g_{c}$\qquad}&
\textrm{$\omega$}&
\textrm{$\chi^{2}/d.o.f$}\\

\colrule
$R_{2}$     & 0.548532(6)  & 1.14(2)  & 0.89\\
$\chi_{u}L$  & 0.548522(8)  & 0.83(4)  & 0.88\\
$\rho_{s}L$  & 0.548521(5)  & 0.68(2)  & 1.09\\

\colrule
$R_{2}$     & 0.456985(6) & 1.08(2)  & 0.63\\
$\chi_{u}L$  & 0.456972(8) & 0.86(3)  & 0.67\\
$\rho_{s}^{x}L$  & 0.456975(5) & 0.70(2)  & 0.92\\
$\rho_{s}^{y}L$  & 0.456983(6) & 0.60(3)  & 0.95\\

\colrule
$R_{2}$     & 0.31446(1)    & 1.08(5)  & 1.11\\
$\chi_{u}L$  & 0.314441(9)   & 0.87(5)  & 0.80\\
$\rho_{s}L$  & 0.314449(6)   & 0.67(3)  & 0.92\\
\end{tabular}
\end{ruledtabular}
\label{tab1}
\end{table}

In order to obtain a better estimation of the critical points, we continue to deal with the crossing points by joint fits as all size dependencies of $g_{c}(L)$ for different quantities should converge to the same value in one system. Therefore we fix $g_{c}(\infty)$ to be the same in each curve and fit all data together with other parameters being independent and $1/\nu=1.406$. The fitting results are shown in all curves in Fig.\ref{fig3} with $g_{c}=0.548524(3)$ in C$2\times2$, $0.456978(2)$ in C$2\times 4$, and $0.314451(3)$ in C$4\times4$. Our result for the C$2\times2$ model is fully consistent with the value in Ref.~[\onlinecite{wenzel09}] with higher precision. By comparing these critical point values we find $g_{c}$ gets smaller from the C$2\times 2$ model to the C$4\times 4$ model, indicating that our models more easily turn into the QPM state with less strong couplings in the unit cell. Therefore we deduce it is a universal rule in QPTs of any C$L_{x}\times L_{y}$ models.

From the separately fitting results in Table~\ref{tab1}, we find that with only one correction term included the correction exponents $\omega$ are not the same for different quantities in the same model, while they are the same for same quantity in different models within at most two error bars with taking average of  $\rho_{s}^{x}$ and $\rho_{s}^{y}$ in the C$2\times4$ model. The difference shows that $\omega$ calculated here is more likely  to be an ``effective correction'' including higher orders. However, fitting including $2\omega$ or higher order is very difficult and challenging with too many free parameters. Here we did not find any nonmonotonic behavior in the size dependence of all crossings in the plaquette models as shown in Fig.~\ref{fig3}; therefore, one correction term can also give convincing criticality analysis, which is also confirmed by the $\chi^{2}/d.o.f$ of each fit. The joint fitting results still share the same rule as the separate ones. Thus, we can estimate the effective $\omega$ by taking the weighted average values of the joint fitting results of $\omega$ from three models. We have the effective correction exponent $\omega=1.058(7)$ for $R_{2}$, for $\chi_{u}L$ $\omega=0.834(7)$. For $\rho_{s}L$ we first get the average $\omega$ from the correlated results of $\rho_{s}^{x}$ and $\rho_{s}^{y}$ in the C$2\times4$ model and we take the larger error of them as the error, which gives $\omega=0.65(1)$ in the end. Then taking the weighted average of all three values gives $\omega=0.66(1)$ for $\rho_{s}L$. Comparing with the standard correction exponent $\omega\approx0.78$\cite{classical} in the O(3) universality class, we can see that the system sizes included in our fits are still not large enough to rule out the affection of higher order corrections even with $L$ up to $160$. Therefore the value of the effective $\omega$ becomes very important in the FSS study to obtain the critical point and critical exponents.

\subsection{Universal quantities at critical points}
As discussed above, in order to study the critical point we use the fixed standard O(3) value $1/\nu=1.406$. The goodness of all fitting results also proves the theoretical prediction. In this section we consider further tests by studying some universal properties to give further evidence of the universality class of the phase transitions.

\begin{figure}
  \includegraphics{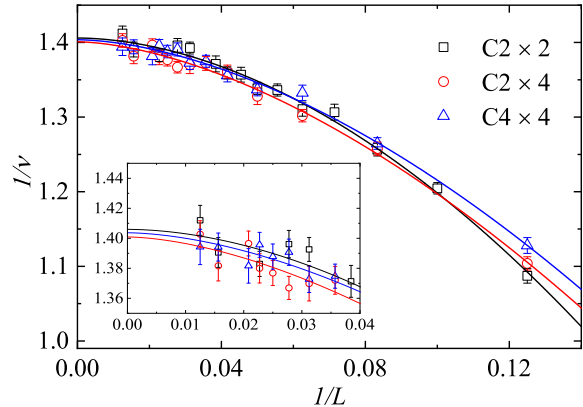}\\
  \caption{(Color online) The size dependence of  $1/\nu(L)$ defined in Eq.(\ref{defnu}) for all models. All data are fitted with Eq.~(\ref{scalenu}) and give $1/\nu=1.406(6)$ in C$2\times2$, $1/\nu=1.401(6)$ in C$2\times4$,
  and  $1/\nu=1.404(5)$ in C$4\times4$. The correction exponents $\omega$ in Eq.~(\ref{scalenu}) are $1.8(1)$, $1.6(1)$, and $1.7(1)$ for C$2\times2$,  C$2\times4$, and C$4\times4$ respectively. The $\chi^{2}/d.o.f$ of all
  fittings are close to 1. The inset figure zooms in with the same data and fitted curves for only larger system sizes to show details of the convergence more clearly.}
 \label{fig4}
\end{figure}

We start with two critical exponents $\nu$ and $\eta$, which are two sensitive universal quantities derived from the Binder ratio and magnetization at the critical point, respectively, and share scaling behavior similar to that of the physical quantities as we discussed in Sec.~\ref{sec:2}. To begin with, the correlation length exponent is calculated using the scaling of $1/\nu(L)$, which is defined as Eq.~(\ref{defnu}), in Eq.~(\ref{scalenu}).  The simulation and scaling results are shown in Fig.~\ref{fig4}
for all three models.  Fitting values of $1/\nu$  are the same for all models considering error.  The weighted average of all three $1/\nu$ value is $1.404(4)$. Compared with the best estimate of $1/\nu=1.4061(7)$ (reciprocal
value of $\nu=0.7112(5)$ in Ref.~[\onlinecite{classical}]) in O(3) it is proved again that the QPTs here are in the same universality class as the CDM, the SDM, and the 3D classical Heisenberg model. However, the accuracy of the estimation
using scaling of $1/\nu(L)$ in our work is much less compared to the previous results. Usually $\nu$ can be obtained from the data collapse together with the critical point and corrections. Here we use the combination of two sizes
together at once in order to lease the influence of the corrections. It does help as the fitting results of $\omega$ are very large in our study, which means that $1/\nu$ converges very fast with the increase of $L$. But
the values of $1/\nu(L)$ obtained from simulations have much larger errors compared with other quantities studied before, which brings in larger error to the final extrapolation value. Much more computational effort is needed
in order to obtain better estimation of $\nu$. We just stop here in this paper as it is not a key point of our work, but we want to point it out for other studies using this procedure.

\begin{figure}
  \includegraphics{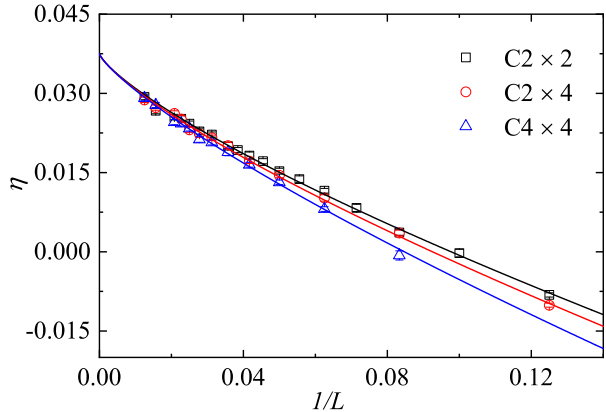}\\
  \caption{(Color online) The size dependence of  $\eta(L)$ defined in Eq.(\ref{defineta}) for all three models. The function used for fitting curves connecting simulation data is Eq.~(\ref{etascale}) with $\eta=0.0375$
  fixed from the 3D classical Heisenberg model, and $\omega=0.77(1)$ in both the C$2\times2$ and C$2\times4$ models with $\chi^{2}/d.o.f\approx1.14$ and $1.16$, while $\omega=0.79(1)$ in the C$4\times4$ model with
  $\chi^{2}/d.o.f\approx0.72$.}
 \label{fig5}
\end{figure}

Another critical exponent considered here is the anomalous dimension $\eta$. Once the critical point $g_{c}$ is obtained, we can study the scaling of order parameter at $g_{c}$ using
 \begin{equation}
\langle m_{s}^{2}\rangle\propto L^{-(1+\eta)}(1+aL^{-\omega}).
\label{mm}
\end{equation}
Similar to $1/\nu(L)$, we can also define $\eta(L)$ from the scaling of pairs of size $L$ and $2L$ as
 \begin{equation}
\eta(L)=\frac{ln[\langle m_{s}^{2}(L)\rangle/\langle m_{s}^{2}(2L)\rangle]}{ln(2)} -1.
\label{defineta}
\end{equation}
In this way, the size dependence of $\eta(L)$ is

 \begin{equation}
\eta(L)=\eta + dL^{-\omega},
\label{etascale}
\end{equation}
with correction to the first order. This time our fits use the best-known estimation of $\eta=0.0375(5)$\cite{classical} and leave the other parameters in Eq.~(\ref{etascale}) free. The fitting results shown in
Fig.~\ref{fig5} again imply that it is correct to set $\eta=0.0375(5)$ here as all QPTs are in the O(3) universality class. Furthermore, all corrections are the same in the three models considering error and the weighted
averaged $\omega=0.78(1)$ is the same as the first correction exponent $\omega_{1}=0.782(13)$ in the O(3) model\cite{classical}. This shows that $\eta$ could be a good quantity in testing the correction exponent in the FSS
study once a good estimation of $g_{c}$ is obtained. And these scaling results also give us more confidence in the accuracy of $g_{c}$  here for all three plaquette models.

\begin{figure}
  \includegraphics{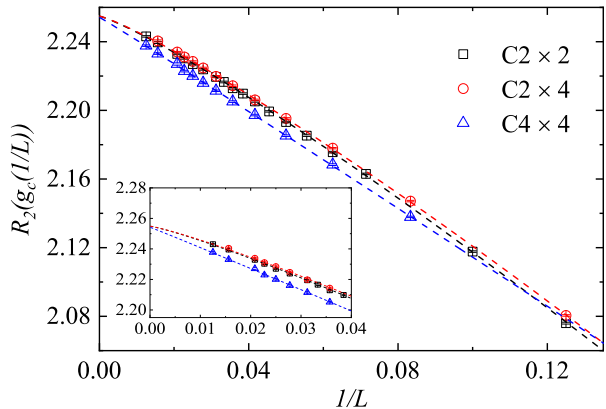}\\
  \caption{(Color online) The Binder ratio $R_{2c}$ at all crossings of $(L,2L)$ versus $1/L$ for C$2\times2$, C$2\times4$, and C$4\times4$ models. All data points are connected by fitting curves using Eq.~(\ref{scaling})
  with the parameters in Table~\ref{tab2}.}
  \label{fig6}
\end{figure}

At last we test the universal quantity Binder ratio $R_{2}$ in the thermodynamic limit. With crossing points extracted from two different sizes $(L and 2L)$ of the $g$ dependence for $R_{2}$ near the critical point, we can obtain the scaling of $R_{2c}(L)$ in Fig.~\ref{fig6}.  The results of fits using Eq.~(\ref{scaling}) in Table~\ref{tab2} indicate that $R_{2c}$ converges to the same value within one error bar in three models with different checkerboard patterns. Except for a further proof of the same universality class, the same $R_{2c}$ in all cases implies that different plaquette models might have the same aspect ratios as well. Taking the weighted average of all $R_{2c}$ values gives $R_{2c}=2.2547(4)$ for a series of plaquette models. Thus, we predict that in all C$L_{x}\times L_{y}$ models according to our model definition, the Binder ratios $R_{2}$ would all converge to $2.2547(4)$ as long as there is an even number of spins in one unit cell with simulated $\beta=L$.

\begin{table}[h]
\caption{\label{tab2}
Estimate results of the critical binder ratio $R_{2c}$ for the C$2\times2$, C$2\times 4$, and C$4\times4$ models. The fitted curves are shown in Fig.~\ref{fig6}.
}
\begin{ruledtabular}
\begin{tabular}{ldcc}
\textrm{}&
\multicolumn{1}{c}\textrm{$R_{2c}$\qquad}&
\textrm{$\omega$}&
\textrm{$\chi^{2}/d.o.f$}\\
\colrule
C$2\times2$            & 2.2549(5)  & 1.163(9)  & 1.00\\
C$2\times4$  	       & 2.2549(7)  & 1.17(1)  & 1.29\\
C$4\times4$            & 2.2542(9)  & 1.02(2)  & 0.84\\
\end{tabular}
\end{ruledtabular}
\end{table}

\section{Summary and discussions}\label{sec:4}
In this paper we carried out the FSS study on data with high-precision using the SSE QMC method. The criticality of three $S=1/2$ Heisenberg models on the square lattice with strong and weak couplings in
plaquette patterns C$2\times2$, C$2\times4$, and C$4\times4$ was studied using the Binder ratio, uniform susceptibility and spin stiffness. By the joint fits combining the scalings of crossing points from all three quantities
we have obtained the most accurate estimates of the critical points $g_{c}$ for three plaquette models up to now. Our scaling analysis implies the importance of corrections in FSS, and with only one correction term the value of
$\omega$ is more likely to be an effective one. The effective $\omega$ does not change in different models as long as it describes the scaling behavior of the same physical observable.  The calculation of $1/\nu$, $\eta$
and $R_{2}$ at the critical point shows that QPTs in all three models are in the O(3) universality class as predicted. The scaling of $\eta$ using the order parameter at the critical point also gives $\omega\approx0.78$, the same as
$\omega_{1}$ determined in the 3D classical Heisenberg model, which further supports the estimate of the critical points.

The fitting results of $\omega$ using different quantities in these three models help us to understand the influence of corrections in the scalings. With system sizes up to $L=160$ the correction exponent $\omega$ is still
an effective one that differs with different variables. However, fitting including higher orders of correction terms would be quite challenging and difficult. Here we find that for models with detailed difference structures in our
case the effective $\omega$ does not change for the same quantity. This might be helpful in further FSS studies on other similar models.  We also obtain the value of the universal quantity $R_{2}$ at the critical point, and we suggest that any C$L_{x}\times L_{y}$ models might have the same critical Binder ratio value with $L_{x}L_{y}$ being even and the same $\beta/L$. This value could be a very important referee in later study of quantum phase transitions, since the same Binder ratio value could be very convincing supporting evidence to show whether a new phase transition belongs to the O(3) universality class.

\begin{acknowledgments}
We thank A. W. Sandvik for useful discussions and a careful reading of the manuscript. The work of X.R and D.X.Y was supported by Grants No. NKRDPC-2018YFA0306001, No. NKRDPC-2017YFA0206203, No. NSFC-11574404, and No. NSFG-2015A030313176; the National Supercomputer Center in Guangzhou; and the Leading Talent Program of Guangdong Special Projects.
\end{acknowledgments}


\end{document}